# Pulsars and Redshifts

David Apsel[1]

## ABSTRACT

**Gravitational redshifts of neutron stars have a theoretical upper limit of z=0.62. Also, it is generally believed that neutron stars have magnetic fields of $10^{12}$-$10^{13}$ G. A previously predicted electromagnetic time dilation effect has been shown to correctly predict decay lifetimes of muons bound to atomic nuclei. In this paper it is shown that the electromagnetic time dilation effect, along with the gravitational time dilation effect, can produce total neutron star redshifts that are substantially larger than 0.62. For instance, the redshift can cutoff radiation for B~$10^{13}$ G. Consequently, we can have a neutron star that is unobservable except for localized surface regions where the cutoff condition does not hold. Assuming coherent radiation, a surface region of this kind that does not include the star's rotation axis will emit a lighthouse type beam. Since the magnetic field in these regions will usually be strong enough to cause significant redshifts, and there is no reason to expect these regions to always be of constant size, shape, or field strength, this model explains the predominance of radio waves, and the existence of pulse variations (e.g. nulling and drifting) in pulsars.**

## INTRODUCTION

It is generally believed that neutron stars have magnetic fields in the range of $10^{12}$-$10^{13}$ G. If appropriate conditions on the equation of state and on stability are applied, it has been shown by Bondi[1] that gravitational redshifts have an upper limit of z=0.62. For pulsars, the pulse is formed by a beam of radiation that sweeps across the observer's line of sight, as in a lighthouse.[2] In previous papers,[3,4] an investigation of the Bohm Aharonov experiment led to the prediction of an electromagnetic time dilation effect analogous to the gravitational time dilation effect. The electromagnetic time dilation effect has been shown to correctly predict decay lifetimes of muons bound to atomic nuclei.[5] This effect has also been discussed at by Rodrigues, et al.,[6] Beil,[7] and Krori, et al.[8] In the current paper it is shown that for a standard neutron star model the electromagnetic time dilation effect, along with the gravitational time dilation effect, can produce total redshifts that are substantially larger than 0.62. For instance, the redshift can cutoff radiation for B~$10^{13}$ G. Consequently, we can have a neutron star that is unobservable except for localized surface regions where the cutoff condition does not hold. Assuming coherent radiation, a surface region of this kind that does not include the star's rotation axis will emit a lighthouse type beam. Since the magnetic field in these regions will usually be strong enough to cause significant redshifts, and there is no reason to expect these regions to always be of constant size, shape, or field strength, this model explains the predominance of radio waves, and the existence of pulse variations (e.g. nulling and drifting) in pulsars.[2,9]

---
[1] 925 Andover Terrace, Ridgewood, NJ.

**2. REDSHIFTS**

Let $g_{\mu\nu}$ and $A_\mu$ be the gravitational and electromagnetic potentials at the surface of a neutron star. From a previous paper,[3] the proper time interval for a displacement $dx^\mu$ at the surface of the star is

$$d\tau = \frac{1}{c}[(g_{\mu\nu}dx^\mu dx^\nu)^{\frac{1}{2}} + \frac{\rho'}{\rho c^2}A_\mu dx^\mu]$$
$$= \frac{1}{c}(g_{\mu\nu}dx^\mu dx^\nu)^{\frac{1}{2}} + \frac{1}{\rho c^2}[\rho' A_4 - \frac{1}{c}\underset{\sim}{J}\cdot \underset{\sim}{A}]dt \quad (1)$$

where $\rho'$ is the scalar charge density, $\rho$ is the scalar mass density, $\underset{\sim}{J}$ is the 3-current density, c is the speed of light, and $\underset{\sim}{A}$ is the 3 vector potential. For a first approximation it can be assumed that $\rho' A_4 = 0$, and that $g_{\mu\nu}$, $\rho$, $\underset{\sim}{J}$, and $\underset{\sim}{A}$ are independent of time. But $\frac{1}{2c}\underset{\sim}{J}\cdot\underset{\sim}{A}$ is the magnetic energy density. The magnetic energy density also has the form $\frac{B^2}{8\pi}$, where B is the magnitude of the magnetic field. Thus we can write

$$d\tau = \frac{1}{c}(g_{\mu\nu}dx^\mu dx^\nu)^{\frac{1}{2}} - \frac{B^2}{4\pi\rho c^2}dt. \quad (2)$$

For an oscillator, at rest, emitting at the surface of a neutron star

$$\frac{\nu}{\nu_0} = (g_{44})^{\frac{1}{2}} - \frac{B^2}{4\pi\rho c^2} \quad (3)$$

where $\nu$ is the frequency observed at infinity and $\nu_0$ is the proper frequency. Consequently,

$$z \equiv \frac{\lambda - \lambda_0}{\lambda_0} = [(g_{44})^{\frac{1}{2}} - \frac{B^2}{4\pi\rho c^2}]^{-1} - 1. \quad (4)$$

Let us define the magnetic and gravitational redshifts as

$$z_m = (1 - \frac{B^2}{4\pi\rho c^2})^{-1} - 1$$
$$z_g = (g_{44})^{-\frac{1}{2}} - 1. \quad (5)$$

Then

$$(z+1)^{-1} = (z_g+1)^{-1} + (z_m+1)^{-1} - 1. \quad (6)$$

For our sun, $z_g \sim 10^{-6}$ and $z_m \sim 10^{-13}$.

For matter with atomic number N, with

$$1 << \eta \equiv (\frac{B}{4.6\times 10^9 N^3})^{\frac{1}{2}} \quad (7)$$

and with B in the range $10^{12}$-$10^{13}$ G, the density has been estimated[10,11] to be

$$\rho \sim 10^6 \left(\frac{N}{26}\right)^4 \eta^{\frac{12}{5}}. \quad (8)$$

Putting this with the expression for z, we can solve for B

$$B \sim N^{\frac{1}{2}}[6.7 \times 10^{10}\{(g_{44})^{\frac{1}{2}} - (z+1)^{-1}\}]^{\frac{5}{4}}. \quad (9)$$

For z >>1, $z_g$ ~ 0.6, and N = 1 or N=4 we get B ~ $10^{13}$ G.

Consequently, we can have a neutron star that is unobservable except for localized surface regions where the cutoff condition does not hold. Assuming coherent radiation, a surface region of this kind that does not include the star's rotation axis will emit a lighthouse type beam. Since the magnetic field in these regions will usually be strong enough to cause significant redshifts, and there is no reason to expect these regions to always be of constant size, shape, or field strength, this model explains the predominance of radio waves, and the existence of pulse variations (e.g. nulling and drifting) in pulsars.[2,9]